\documentclass[11pt,letterpaper]{article}
\usepackage[letterpaper,margin=1in]{geometry}
\usepackage{amsmath}
\usepackage{amssymb}
\usepackage{graphicx}
\usepackage{booktabs}
\usepackage{microtype}
\usepackage[authoryear,round]{natbib}
\usepackage[hidelinks]{hyperref}
\graphicspath{{./}{figures/}}

\newcommand{\Prob}{\mathrm{P}}
\newcommand{\Expect}{\mathrm{E}}

\title{Reconciling District Prediction Markets and Seat-Count Buckets in U.S. Midterm Elections}

\author{
Michael Tan\thanks{Union County Magnet High School, Scotch Plains, NJ 07076. Email: \texttt{mtan@ucvts.org}.}
\and
Han Xiao\thanks{Department of Statistics, Rutgers University, Piscataway, NJ 08854. Email: \texttt{hxiao@stat.rutgers.edu}.}
}

\date{}

\begin{document}
\maketitle

\vspace{-2em}
\begin{abstract}
The prediction market platform Kalshi offers individual markets for the winner in each U.S. congressional district and an aggregate market for the total number of Democratic House seats. We develop an applied statistical framework for reconciling these two market layers, since the aggregate seat count is determined by the collection of individual district outcomes. The framework combines a latent-Gaussian dependence structure with a log-odds adjustment to district probabilities and a power adjustment to aggregate seat-count probabilities, with parameters estimated by divergence minimization. 
Applied to a July 30, 2026 snapshot, the three-parameter primary model achieves close reconciliation, with forward and reverse KL divergences both approximately 0.0013, while yielding a positive latent national correlation, a negative district log-odds shift, and flattening of the aggregate target distribution.
Daily analysis of 236 dates from January through August 2026 reveals two relatively stable periods separated by a spring transition period exhibiting substantial volatility in both parameter and divergence values. The framework provides a diagnostic approach for assessing cross-market consistency in hierarchically structured prediction or financial markets.
\end{abstract}



\section{Introduction}
\label{sec:introduction}

Prediction markets such as Kalshi and Polymarket are trading platforms where participants buy and sell contracts based on the outcomes of future events. The market prices of these contracts are often interpreted as collective probabilistic forecasts. These markets can represent the same underlying outcomes at different levels of aggregation. In the context of U.S. congressional elections, for example, an individual district contract concerns the winner of a single congressional race, while the aggregate seat-count market concerns the total number of seats won nationally by a political party. Prediction markets are widely studied as mechanisms for aggregating dispersed information \citep{wolfers2004prediction}. However, prices across these two types of markets need not exhibit probabilistic consistency, even though the underlying contracts depend on the same collection of district election results.

Motivated by discrepancies between district-level and aggregate seat-count market probabilities, we present a modeling framework to reconcile these two market layers and study how their relationship evolves over time. The purpose is not to estimate who will win the 2026 midterm elections or to present a new forecast. Rather, these different types of election contracts provide a setting for studying consistency across related markets. Two core issues arise. First, individual district markets provide marginal probabilities but no information about dependence across districts. Second, introducing a dependence structure alone does not fully reconcile the district-level probabilities with the aggregate market; structural adjustments to the price-derived probabilities are also needed.

Our framework introduces a correlated latent-Gaussian structure, a log-odds adjustment to the input district probabilities, and a power adjustment to the aggregate seat-count probabilities. We fit the model by minimizing a divergence measure between the model-implied aggregate distribution, derived from the district inputs and latent dependence structure, and the adjusted aggregate market distribution. Using publicly available Kalshi data from January 8 through August 31, 2026, we apply and evaluate the framework in two main parts. Specifically, the \textbf{snapshot analysis} uses the July 30, 2026 snapshot to assess robustness across divergence criteria and explore model extensions, while the \textbf{historical daily analysis} examines how market reconciliation evolves over time, including periods of relative stability and substantial volatility.

From an applied statistical perspective, our modeling framework offers potential practical value as a diagnostic tool for analyzing cross-market consistency. Fitting the model over time allows the estimated parameters and divergence measures to characterize periods of relative alignment or misalignment between individual and aggregate markets. Although this paper focuses on the 2026 U.S. midterm cycle, the methodology can be adapted to other systems of hierarchically related contracts in broader financial markets.

Relatively little research has focused on reconciling individual district prediction markets with aggregate seat-count markets. Our work builds upon and connects several distinct strands of literature. Prediction markets are widely recognized for their capacity to aggregate dispersed information \citep{wolfers2004prediction}, while ensuring structural consistency across related contracts remains challenging \citep{chen2010designing}. To capture dependence among individual events, we use a latent-variable framework related to Bayesian probit modeling for binary data \citep{albert1993bayesian} and Gaussian copula methods in quantitative finance \citep{cherubini2004copula}. The use of structural adjustments to raw market probabilities is motivated by the behavioral finance literature, particularly evidence on the longshot bias in parimutuel and prediction markets \citep{thaler1988parimutuel,snowberg2010explaining,page2013do}.


\section{Data}
\label{sec:data}

This paper examines two types of prediction markets on Kalshi for the 2026 U.S. midterm elections: 435 individual district markets (e.g., \citealp{KalshiNJ7} for New Jersey's 7th congressional district) and twelve aggregate seat-count buckets \citep{KalshiHouseSeats}.
For the district markets, this paper uses the Democratic side of each binary market, yielding one market-price input per district. For the aggregate market, it uses the YES side of each Democratic seat-count bucket, yielding one input probability per bucket.
The twelve seat-count buckets partition all possible Democratic seat totals from zero to 435. These consist of a lower tail bucket for totals below 210, ten bounded intermediate buckets containing four consecutive seat totals each ([210, 213], [214, 217], \dots, [246, 249]), and an upper tail bucket for totals above 249.

\subsection{Source and Processing}
\label{subsec:source-processing}

The snapshot data were collected using Kalshi's order book API. For binary markets, the order book API returns YES and NO bids rather than asks. A NO bid at price $b^\mathrm{NO}$ implies a YES ask at price $1-b^\mathrm{NO}$, with prices in dollars; YES asks can therefore be reconstructed from the NO bids \citep{kalshiorderbook}. The saved snapshot files carry timestamps around 4:00 p.m. Eastern Time (ET) on July 30, 2026. District order books were requested sequentially, and thus the data do not constitute an exactly simultaneous snapshot.

Historical data were obtained from Predexon, which maintains tick-level archived Kalshi order books \citep{predexon}. The target observation time was 4:00 p.m. ET, and archived snapshots were selected by proximity to that target. Historical observations were initially selected from the nearest archived order book within 30 minutes of the intended observation time. Where snapshots in this range were unavailable, a broader search selected the nearest usable observation. The final selected snapshot could precede or follow the intended observation time, including falling on another calendar date. For illiquid district observations, we carried forward the most recent earlier acceptable price for that market. If no such observation was available, the raw price was retained and the observation was flagged.
The saved data cover 236 dates from January 8, the beginning of the available historical data, through August 31, 2026. This yields a total of 102,660 district observations and 2,832 bucket observations.

The July 30 snapshot data include deterministic inputs for nine California districts with same-party general elections. These California primaries resulted in general-election races between candidates from the same party. For the historical data, these district inputs were fixed after their contracts were finalized, whereas prior dates retain active market prices.

\subsection{Converting to Probabilities}
\label{subsec:probability}

In binary prediction markets where a winning contract pays out exactly one dollar, the trading price of a contract is commonly interpreted as the market-implied probability of the event occurring. For district $i$, let $b_i$ and $a_i$ be the best YES bid and best YES ask, and let $v_i^b$ and $v_i^a$ be the best-bid and best-ask quantities, respectively. To estimate the probability of a Democratic win in district $i$, we use the microprice defined as \citep{CarteaJaimungalPenalva2015}
\begin{equation}
    \pi^0_i = \frac{a_i v_i^b + b_i v_i^a}{v_i^a + v_i^b}.
    \label{eq:microprice}
\end{equation}
The microprice weights the top-of-book bid and ask prices by the quantities available at the opposite side of the order book. A larger bid-size quantity moves this estimator toward the ask, while a larger ask-size quantity moves it toward the bid. This provides a more informative price-derived probability than simply using the bid-ask midpoint. The same microprice formulation is applied to obtain the probabilities for the seat-count buckets.

\begin{figure}[tbhp]
    \centering
    \includegraphics[width=.8\textwidth]{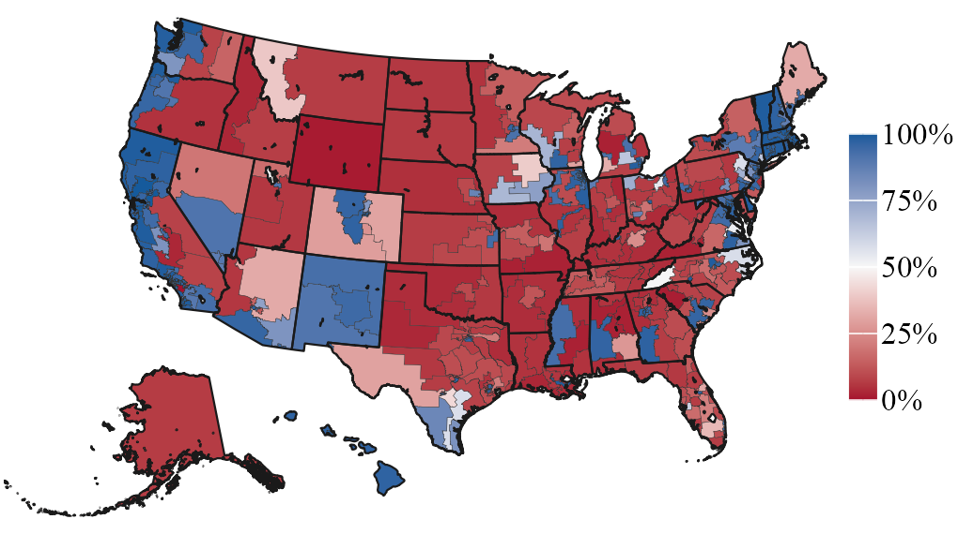}
    \caption{Map of the Democratic contract microprices representing the market-implied probabilities of Democratic wins across the 435 districts, using data from the July 30, 2026 snapshot. Blue indicates higher probabilities and red indicates lower probabilities.}
    \label{fig:microprices}
\end{figure}

Because the district markets correspond to separate binary events rather than mutually exclusive outcomes, their microprices represent marginal probabilities and are used without normalization across districts.
Figure~\ref{fig:microprices} displays these price-derived probabilities across the 435 congressional districts for the July 30, 2026 snapshot.
In contrast, the seat-count buckets are mutually exclusive and exhaustive, so their microprices $\pi_b^0$ are normalized as $p_b^0=\pi_b^0/\sum_{c=1}^B\pi_c^0$, $b=1,\ldots,B$ ($B=12$). The saved snapshot bucket microprices sum to 1.016526 before normalization.

We refer to the above price-derived probabilities, $\pi^0_i$ and $p^0_b$, as initial or input probabilities, which can be further adjusted in the model developed in the next section.


\section{Model}
\label{sec:model}

The model provides a probabilistic framework that connects individual district markets with the aggregate seat-count market. Let $W_i$ be an indicator variable taking the value $1$ for a Democratic win in district $i$ and $0$ otherwise, with marginal probability $\pi_i=\Prob(W_i=1)$, and let $S=\sum_{i=1}^{N}W_i$ denote the total number of Democratic seats won, where $N=435$. The initial district probabilities are the price-derived microprices $\pi_i^0$ defined in Equation~\eqref{eq:microprice}. The following subsections introduce an independent baseline, a latent-Gaussian dependence structure, adjustments to the district and seat-count probabilities, and an extension incorporating regional dependence. The resulting seat-count distribution and the formal reconciliation criterion are described in Section~\ref{sec:methods}.

\subsection{Independent Baseline}
\label{subsec:independence}

As a baseline, consider the case where all districts are independent of each other. Then $S$ is Poisson--binomial with $\Expect(S)=\sum_{i=1}^N\pi_i$ and $\operatorname{Var}(S)=\sum_{i=1}^N\pi_i(1-\pi_i)$.
Independence serves as a useful benchmark because it requires no dependence parameters. However, the resulting seat-count distribution is substantially more concentrated than the distribution implied by the aggregate seat-count market, as shown in Figure~\ref{fig:poisson}. More generally,
\[
\operatorname{Var}(S)
=
\sum_{i=1}^{N}\operatorname{Var}(W_i)
+
2\sum_{i<j}\operatorname{Cov}(W_i,W_j).
\]
Under independence, all covariance terms are zero. Positive correlations between district outcomes increase the dispersion of the seat-count distribution.
Thus, while the marginal district probabilities determine the expected seat total, incorporating dependence across districts is necessary to reproduce the broader dispersion observed in the aggregate market.

\subsection{Correlated Latent-Gaussian Model}
\label{subsec:factor-model}

To introduce dependence while preserving the marginal district probabilities, we use a latent-variable framework \citep{albert1993bayesian}. Let $Z_\mathrm{nat}$ be a shared national factor, and let $\varepsilon_i$ be the district-specific disturbances, with $Z_\mathrm{nat},\varepsilon_1,\ldots,\varepsilon_N$ mutually independent standard Gaussian variables. We define a continuous latent score $Y_i$ whose sign determines the district win indicator $W_i$:
\begin{align*}
    Y_i
        &= \mu_i+\sigma_\mathrm{nat}Z_\mathrm{nat}
           +\sqrt{1-\sigma_\mathrm{nat}^2}\,\varepsilon_i,\quad
    W_i
        =\mathbf{1}\{Y_i>0\}.
\end{align*}
Here $\mu_i=\Phi^{-1}(\pi_i)$, with $\Phi^{-1}$ denoting the standard Gaussian quantile function. By construction, $Y_i$ is Gaussian with mean $\mu_i$ and variance $1$, and $\Prob(W_i=1)=\Prob(Y_i>0)=\Phi(\mu_i)=\pi_i$. Thus the latent dependence structure preserves the marginal district probabilities.

The parameter $\sigma_\mathrm{nat}^2\in[0,1)$ represents the fraction of unit latent variance allocated to the shared national factor.
It is also the correlation between the latent scores of any two distinct districts:
\begin{equation}
    \operatorname{Corr}(Y_i,Y_j)
    =\sigma_\mathrm{nat}^2,
    \qquad i\ne j.
    \label{eq:national-correlation}
\end{equation}
Increasing $\sigma_\mathrm{nat}^2$ shifts latent variance from the district-specific disturbances to the common national factor, thereby increasing dependence while preserving the marginal probabilities $\pi_i$ and the expected seat total $\Expect(S)$.

Conditional on a realization of the national factor, $Z_\mathrm{nat}=z$, the district outcomes become independent, with conditional win probability
\begin{equation}
    \pi_i(z)
    =
    \Phi\!\left(
        \frac{\mu_i+\sigma_\mathrm{nat}z}
             {\sqrt{1-\sigma_\mathrm{nat}^2}}
    \right).
    \label{eq:conditional-input}
\end{equation}
This conditional representation is used in the numerical evaluation of the seat-count distribution described in Section~\ref{sec:methods}.

\subsection{District Probability Adjustment}
\label{subsec:districts}

The national factor changes the dependence among district outcomes and, consequently, the dispersion of the seat-count distribution, but does not alter the marginal district probabilities or the expected seat total $\Expect(S)$. To allow for discrepancies in the location of the district-implied and aggregate seat-count distributions, we apply a log-odds transformation to the initial price-derived district probabilities $\pi_i^0$:
\begin{equation}
    \operatorname{logit}(\pi_i)
    =
    \alpha+\beta\operatorname{logit}(\pi_i^0),
    \qquad 0<\pi_i^0<1,
    \label{eq:district-adjustment}
\end{equation}
where $\operatorname{logit}(x)=\log\{x/(1-x)\}$. The model then sets the district mean score $\mu_i=\Phi^{-1}(\pi_i)$ as in Section~\ref{subsec:factor-model}. Input probabilities $\pi_i^0$ equal to zero or one remain fixed and bypass this adjustment.

The transformation in Equation~\eqref{eq:district-adjustment} is related to Platt scaling \citep{platt2000}. The parameter $\alpha\in\mathbb{R}$ acts as an additive shift in district log-odds: positive values increase all nondegenerate district log-odds, and hence their probabilities, while negative values decrease them.
The parameter $\beta>0$ rescales differences in log-odds across districts. Values of $\beta<1$ compress these differences, values of $\beta>1$ expand them, and $\beta=1$ preserves the original differences in log-odds.

To maintain model parsimony, given that freeing $\beta$ yields only modest reductions in divergence (see Table~\ref{tab:model-comparison}), our primary specification fixes $\beta=1$ and uses $\alpha$ alone to adjust the location of the district probabilities.

\subsection{Bucket Probability Adjustment}
\label{subsec:buckets}

To align the seat-count distribution from the aggregate market with that implied by the district-level model,
we apply a power transformation to the initial price-derived bucket probabilities $p_b^0$, yielding the adjusted bucket probabilities $p_b(\gamma)$:
\begin{equation}
    p_b(\gamma)
    =
    \frac{(p_b^0)^\gamma}
         {\sum_{c=1}^{B}(p_c^0)^\gamma},
    \qquad \gamma>0.
    \label{eq:bucket-adjustment}
\end{equation}
The parameter $\gamma$ controls the relative concentration of the bucket probabilities. Values of $\gamma>1$ increase the relative weight of higher-probability buckets, concentrating mass in those buckets, whereas values of $\gamma<1$ flatten the differences between high- and low-probability buckets. A value of $\gamma=1$ leaves the original bucket probabilities unchanged.

For any two buckets $b$ and $c$, the transformation implies
\begin{equation}
    \log\frac{p_b(\gamma)}{p_c(\gamma)}
    =
    \gamma\log\frac{p_b^0}{p_c^0}.
    \label{eq:gamma-contrast}
\end{equation}
Thus $\gamma$ acts as a common multiplier on every pairwise log-ratio of the original bucket probabilities. In this sense, it is analogous to the district multiplier $\beta$, although $\gamma$ operates on the multi-category aggregate distribution rather than on binary district marginals. The transformation preserves the rank ordering of the bucket probabilities and the unit-sum constraint.

\subsection{Incorporating Regional Correlation}
\label{subsec:region}

To capture geographic dependence beyond the common national factor, we extend the latent model by introducing regional factors. Let $r(i)$ denote the region containing district $i$, and let $Z_{r(i)}$ denote the corresponding standard Gaussian regional factor. The regional factors are mutually independent across regions and are independent of the national factor and district-specific disturbances. The extended latent score is
\begin{equation*}
    Y_i
    =
    \mu_i
    +\sigma_\mathrm{nat}Z_\mathrm{nat}
    +\sigma_\mathrm{rgn}Z_{r(i)}
    +\sqrt{1-\sigma_\mathrm{nat}^2-\sigma_\mathrm{rgn}^2}\,
     \varepsilon_i.
\end{equation*}
Here $\sigma_\mathrm{rgn}^2\ge0$ is the fraction of unit latent variance shared by districts within the same region. The variance components satisfy
$\sigma_\mathrm{nat}^2+\sigma_\mathrm{rgn}^2<1$, ensuring that
$\operatorname{Var}(Y_i)=1$.

The nine regions are the standard U.S. Census divisions: New England, Middle Atlantic, East North Central, West North Central, South Atlantic, East South Central, West South Central, Mountain, and Pacific \citep{censusdivisions}. Districts are assigned to regions according to their state.

Under this specification, the national factor contributes to dependence between every pair of districts, whereas a regional factor is shared only by districts in the same region. Consequently, for two distinct districts, the latent correlation becomes
\begin{equation}
    \operatorname{Corr}(Y_i,Y_j)
    =
    \begin{cases}
        \sigma_\mathrm{nat}^2+\sigma_\mathrm{rgn}^2,
            & \text{if } r(i)=r(j),\\
        \sigma_\mathrm{nat}^2,
            & \text{if } r(i)\ne r(j).
    \end{cases}
    \label{eq:regional-correlation}
\end{equation}
From Equation~\eqref{eq:regional-correlation}, increasing $\sigma_\mathrm{rgn}^2$ increases within-region latent correlation while leaving cross-region correlation unchanged.
To maintain a parsimonious specification, the model uses a single regional variance parameter $\sigma_\mathrm{rgn}^2$ shared across all nine regions rather than allowing a separate variance component for each region.


\section{Methods}
\label{sec:methods}

For a seat-count bucket $b$ with inclusive bounds $L_b$ and $U_b$, let
$q_b(\theta)=\Prob_\theta(L_b\le S\le U_b)$ denote its model-implied probability under the latent-Gaussian district model,
where the parameter vector $\theta$ includes variance components and district adjustment parameters.
The model-implied bucket vector is $q(\theta)=(q_1(\theta),\ldots,q_B(\theta))$. Separately, the adjusted market probabilities defined in Equation~\eqref{eq:bucket-adjustment} form the target vector $p(\gamma)=(p_1(\gamma),\ldots,p_B(\gamma))$.

For the primary specification, $\theta=(\sigma_\mathrm{nat}^2,\alpha)$. The joint parameter vector is therefore
$(\theta,\gamma)=(\sigma_\mathrm{nat}^2,\alpha,\gamma)$, with fixed $\beta=1$ and no regional factor ($\sigma_\mathrm{rgn}^2=0$).
Model extensions introduce a free multiplier $\beta$, regional dependence $\sigma_\mathrm{rgn}^2$, or both.
All model evaluations and comparisons treat the initial price-derived probabilities $\pi_i^0$ and $p_b^0$ as fixed inputs.

\subsection{Fitting Criteria}
\label{subsec:fitting-criteria}

To evaluate alignment between the model-implied bucket vector $q(\theta)$ and the adjusted market vector $p(\gamma)$, we employ four statistical divergence measures \citep{ali1966general}. The forward Kullback--Leibler (KL) divergence, using the market distribution as the reference, is
\[
D_{\mathrm{KL}}\!\left(p(\gamma)\parallel q(\theta)\right)
=
\sum_{b=1}^{B}p_b(\gamma)
\log\!\left(\frac{p_b(\gamma)}{q_b(\theta)}\right).
\]
The reverse KL divergence, using the model distribution as the reference, is $D_{\mathrm{KL}}(q(\theta)\parallel p(\gamma))=\sum_{b=1}^{B}q_b(\theta)\log\{q_b(\theta)/p_b(\gamma)\}$. The Pearson chi-square divergence is $D_{P\chi^2}(p,q)=\sum_{b=1}^{B}\{p_b(\gamma)-q_b(\theta)\}^2/q_b(\theta)$, and the Neyman chi-square divergence is $D_{N\chi^2}(p,q)=\sum_{b=1}^{B}\{p_b(\gamma)-q_b(\theta)\}^2/p_b(\gamma)$.
Agreement among parameter estimates across the different divergence measures provides a sensitivity check on the choice of fitting criterion. A probability floor of $10^{-12}$ is applied to $q_b(\theta)$ and $p_b(\gamma)$ to keep numerical evaluations of logarithms and divisions finite.

For each divergence criterion $D$, parameters are jointly estimated by solving
\[
(\widehat{\theta}_D,\widehat{\gamma}_D)
=
\underset{\theta\in\Theta,\,\gamma\in\Gamma}{\operatorname{arg\,min}}\;
D\!\left(p(\gamma),q(\theta)\right),
\]
where $\Theta$ and $\Gamma$ denote the admissible parameter domains for the latent district model and the bucket transformation exponent, respectively.

For the July 30 snapshot, the primary model is fitted under all four divergence criteria, while the model extensions are fitted under forward KL divergence. For the historical dataset spanning 236 dates, each date is fitted independently using the three-parameter primary model under forward KL divergence as the divergence objective.

\subsection{Numerical Integration and Optimization}
\label{subsec:optimization}

Evaluating $q_b(\theta)$ requires the distribution of
$S=\sum_{i=1}^{N}W_i$. Conditional on the latent factors, the district outcomes are independent Bernoulli variables. Thus the conditional distribution of $S$ is Poisson--binomial and can be evaluated using standard numerical recursions \citep{chen1997}.

For the national dependence specification, $q_b(\theta)$ is computed by integrating the conditional bucket probabilities, using the conditional district probabilities in Equation~\eqref{eq:conditional-input}, over the standard Gaussian national factor $Z_\mathrm{nat}$ via 33-node Gauss--Hermite quadrature \citep{golub1969calculation}.
For the regional dependence specification, for each fixed value of $Z_\mathrm{nat}$, we first compute the seat-count distribution within each region conditional on $Z_\mathrm{nat}$ and that region's regional factor.
We then integrate over each regional factor independently and combine the resulting regional seat-count distributions to obtain the total seat-count distribution conditional on $Z_\mathrm{nat}$. Finally, we integrate over $Z_\mathrm{nat}$ using 33-node Gauss--Hermite quadrature. Deterministic district inputs ($\pi_i^0\in\{0,1\}$) described in Section~\ref{sec:data} remain fixed at zero or one throughout these calculations and bypass the logit and probit transformations.

Parameter estimation uses bounded L-BFGS-B optimization \citep{byrd1995limited}, initialized from three distinct parameter vectors to reduce sensitivity to initial conditions. Optimization bounds are set to
$0.002\le\sigma_\mathrm{nat}^2\le0.45$,
$-1.5\le\alpha\le1.5$, and
$0.3\le\gamma\le3.0$.
When included, the regional variance parameter satisfies
$0.002\le\sigma_\mathrm{rgn}^2\le0.45$ subject to
$\sigma_\mathrm{nat}^2+\sigma_\mathrm{rgn}^2<1$, and free-$\beta$ models use
$0.3\le\beta\le3.0$.
For each fit, the parameter vector yielding the smallest objective value among the converged runs is retained.

Data collection and processing used both Python \citep{Python313} and R \citep{RCoreTeam2026}, while model fitting was performed in R.


\section{Results}
\label{sec:results}

The empirical analysis consists of two parts. The July 30 snapshot analysis examines how alternative model specifications reconcile discrepancies between the individual district markets and the aggregate seat-count market. The historical analysis of daily observations then examines how the fitted parameters and divergence evolve over time. Both analyses focus on reconciliation of the two markets rather than on specific predictions or validated outcome probabilities.

\subsection{Snapshot Results}
\label{subsec:snapshot}

\begin{figure}[tbp]
    \centering
    \includegraphics[
        width=.8 \textwidth
    ]{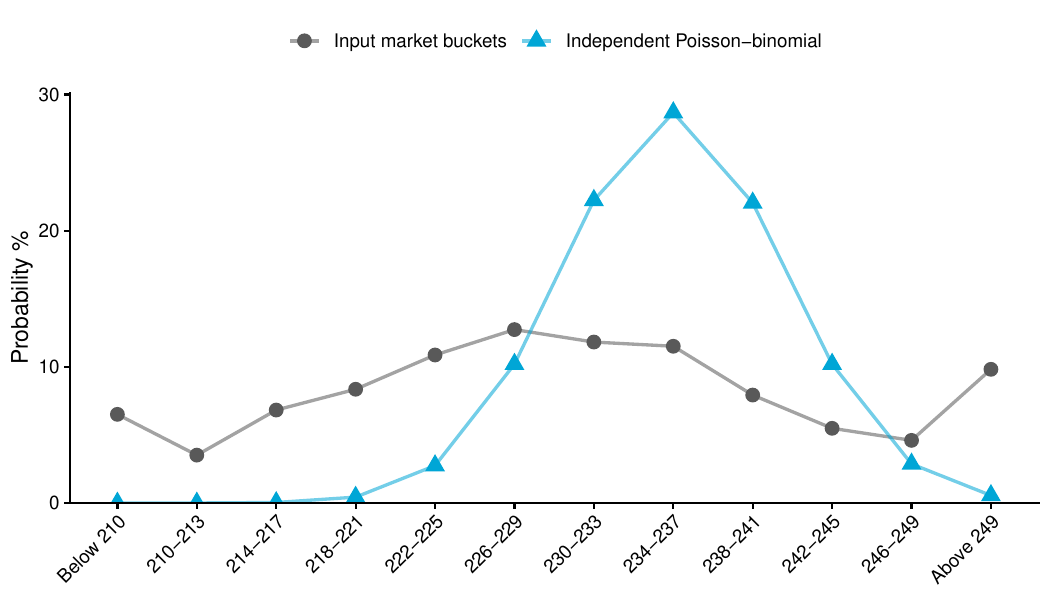}
    \caption{Normalized seat-count market probabilities compared with
    bucket probabilities computed from the Poisson--binomial distribution
    under district independence using the input district microprices. Both
    series use the July 30 snapshot around 4:00 p.m. ET.}
    \label{fig:poisson}
\end{figure}

\textbf{Independence Baseline.}
The independent baseline produces a more concentrated seat-count distribution than the distribution implied by the aggregate market, as shown in Figure~\ref{fig:poisson}. Therefore, aggregating the unadjusted district inputs under independence fails to reproduce the dispersion observed in the aggregate market.

\textbf{Primary Specification.}
Fitting the three-parameter primary model under each of the four divergence criteria yields consistent parameter estimates (Table~\ref{tab:three-parameter-region-fit}). Despite its parsimonious structure, the primary specification achieves a close reconciliation, with minimized forward and reverse KL divergences both approximately $0.0013$. The national variance component $\sigma_\mathrm{nat}^2$ ranges from $0.0576$ to $0.0578$, and the district shift $\alpha$ ranges from $-0.1813$ to $-0.1809$.
All four estimates of the bucket exponent $\gamma$ are below one. This agreement indicates the same qualitative adjustments across the four criteria within the primary specification: a positive latent national correlation, a negative district log-odds shift, and moderate flattening of the seat-count bucket target.

Under forward KL divergence, $D_{\mathrm{KL}}(p\parallel q)$, the estimated national variance component is $\sigma_\mathrm{nat}^2=0.0577$. This assigns $5.77\%$ of each district's latent variance to the shared national factor and, by Equation~\eqref{eq:national-correlation}, indicates a correlation of $0.0577$ between the latent scores of any two districts. For the district inputs, $\alpha=-0.1810$ shifts the log-odds of each nondegenerate district probability downward, lowering the corresponding Democratic probabilities, with the largest absolute probability changes occurring for inputs near $0.5$. Finally, the bucket exponent $\gamma=0.6759$ flattens the target seat-count bucket probabilities, giving relatively more weight to lower-probability buckets.
By Equation~\eqref{eq:gamma-contrast}, it scales every pairwise log-ratio of the original bucket probabilities by $0.6759$, corresponding to a $32.41\%$ contraction in these log-ratios.

\begin{table}[tbp]
    \centering
    \caption{Primary national-only model estimates under alternative divergence criteria}
    \label{tab:three-parameter-region-fit}
    \vspace{0.5em}
    \normalsize
\begin{tabular}{lrrrr}
\toprule
Criterion & $\sigma^2_{\mathrm{nat}}$ & $\alpha$ & $\gamma$ & Divergence\\
\midrule
$D_{\mathrm{KL}}(p \parallel q)$ & 0.0577 & -0.1810 & 0.6759 & 0.001315\\
$D_{\mathrm{KL}}(q \parallel p)$ & 0.0577 & -0.1812 & 0.6726 & 0.001305\\
Pearson $\chi^2$ & 0.0576 & -0.1809 & 0.6792 & 0.002650\\
Neyman $\chi^2$ & 0.0578 & -0.1813 & 0.6693 & 0.002591\\
\bottomrule
\end{tabular}

    \vspace{0.5em}
    \begin{minipage}{0.95\textwidth}
        \small
        \textit{Note:} The parameter $\sigma_\mathrm{nat}^2$ is the national variance component and also the correlation between the latent scores of any two districts; $\alpha$ and $\gamma$ are the district and bucket adjustment parameters, respectively; $p(\gamma)$ is the adjusted market vector and $q(\theta)$ is the model-implied bucket vector; Divergence values are criterion-specific and are not directly comparable across rows.
    \end{minipage} \vspace{-0.5em}
\end{table}

\begin{figure}[tbp]
    \centering
    \includegraphics[
        width=.8 \textwidth
    ]{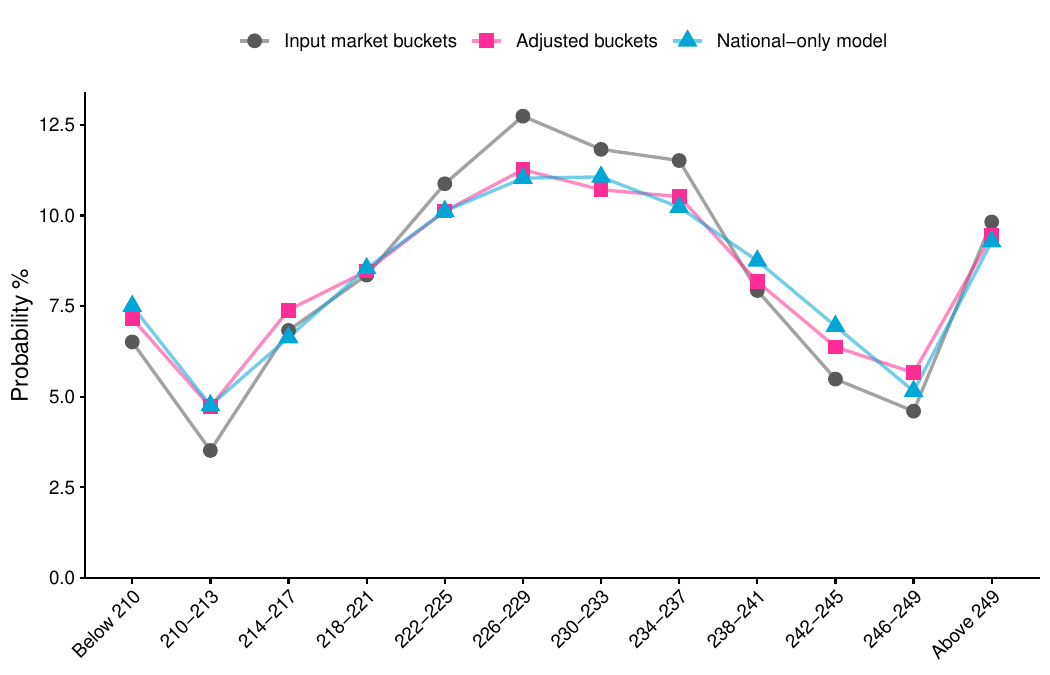} \vspace{-1em}
    \caption{Comparison of normalized seat-count market probabilities,
    their $\gamma$-adjusted counterparts, and the fitted national-only
    primary model using the July 30 snapshot. The model is fitted using forward KL divergence $D_{\mathrm{KL}}(p \parallel q)$.}
    \label{fig:region-fit-residuals}
\end{figure}

For forward KL divergence, Figure~\ref{fig:region-fit-residuals} displays the primary model's fitted bucket probabilities alongside the original and adjusted market bucket probabilities. The fitted probabilities align closely with the adjusted market target.

\textbf{Alternative Specifications.}
Table~\ref{tab:model-comparison} uses forward KL divergence to compare the primary model with alternative specifications. These comparisons add regional correlation, district log-odds rescaling through a free multiplier $\beta$, or both.

\begin{table}[tbp]
    \centering
    \caption{Comparison of alternative model specifications under forward KL divergence.}
    \label{tab:model-comparison}

    \vspace{0.5em}
    \normalsize
\begin{tabular}{lrrrrrr}
\toprule
Model & $\sigma^2_{\mathrm{nat}}$ & $\alpha$ & $\gamma$ & $\beta$ & $\sigma^2_{\mathrm{rgn}}$ & $D_{\mathrm{KL}}(p \parallel q)$\\
\midrule
$\sigma^2_{\mathrm{nat}},\ \alpha,\ \gamma$
    & 0.0577 & -0.1810 & 0.6759 & -- & -- & 0.001315\\
$(+\,\beta)$
    & 0.1184 & -0.1580 & 0.7615 & 1.2241 & -- & 0.001143\\
$(+\,\sigma^2_{\mathrm{rgn}})$
    & 0.0133 & -0.1775 & 0.7262 & -- & 0.3174 & 0.001113\\
$(+\,\beta\ \mathrm{and}\ \sigma^2_{\mathrm{rgn}})$
    & 0.0133 & -0.1775 & 0.7262 & 0.9999 & 0.3174 & 0.001113\\
\bottomrule
\end{tabular}

    \vspace{0.5em}
    \begin{minipage}{0.95\textwidth}
        \small
        \textit{Note:} A dash indicates fixed $\beta=1$ or the absence of regional dependence.
    \end{minipage}\vspace{-0.5em}
\end{table}

Adding regional dependence lowers the forward KL divergence to $0.001113$, approximately a $15.3\%$ reduction from the primary model. The national variance component falls from $0.0577$ to $0.0133$, while the regional variance component is estimated as $0.3174$. The estimate of $\alpha$ remains close to that from the primary model, and the estimate of $\gamma$ increases from $0.6759$ to $0.7262$.

Allowing $\beta$ to vary without regional dependence yields a divergence of $0.001143$, approximately $13.1\%$ below that of the primary model. The estimated multiplier is $\beta=1.2241$, which expands differences in the original district log-odds by approximately $22.41\%$.
The estimated latent national correlation increases from $0.0577$ to $0.1184$, the estimate of $\alpha$ remains negative but changes from $-0.1810$ to $-0.1580$, and the estimate of $\gamma$ increases from $0.6759$ to $0.7615$.

When regional dependence and a free $\beta$ are both included, the estimate $\beta=0.9999$ is effectively one, with almost no further reduction in divergence relative to the fixed-$\beta$ regional model.

Overall, the additional parameters produce modest reductions in forward KL divergence. They lead to noticeable changes in the fitted dependence structure, while the district shift and bucket adjustment remain broadly similar to those in the primary model. In particular, freeing $\beta$ changes the degree of adjustment and national dependence in the national-only model but adds little once regional dependence is included.

\subsection{Historical Daily Results}
\label{subsec:historical}

The second part of the empirical analysis studies reconciliation across 236 daily observations from January 8 through August 31, 2026. Each date uses the same specification: the primary model with estimated $\sigma_\mathrm{nat}^2$, $\alpha$, and $\gamma$, fitted using forward KL divergence. All reported daily optimizations converged successfully. District markets treated as deterministic after their cutoff switch from historical market inputs to fixed values as described in Section~\ref{subsec:source-processing}.

Over the eight-month history, the fitted parameters exhibit a period of substantial volatility during the spring, approximately from March 27 to May 8. Outside this window, the parameter estimates fluctuate in a relatively stable manner. The two dates delimit an observed period of volatility and are descriptive rather than estimated change points.

\begin{figure}[tbp]
\centering
\includegraphics[width=\textwidth]{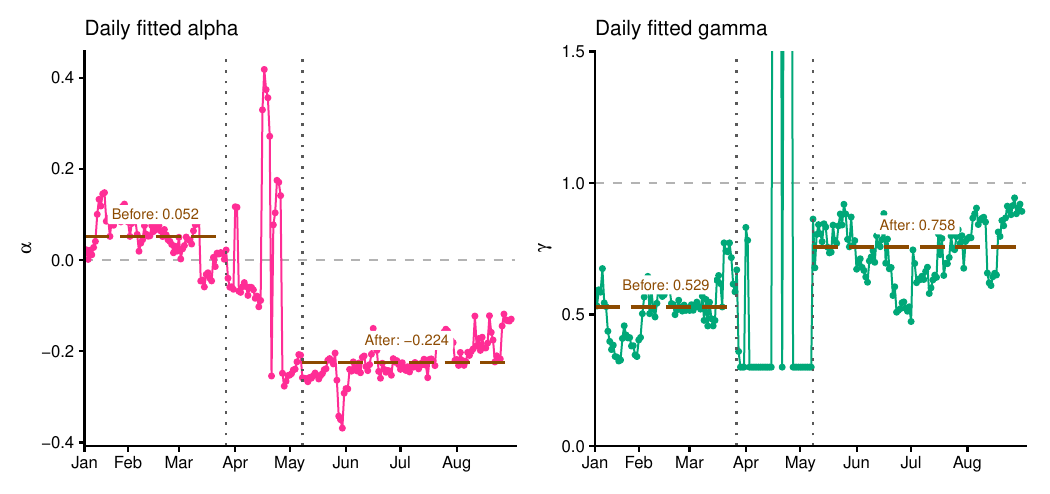}
\caption{Daily fitted district adjustment $\alpha$ (left) and bucket exponent $\gamma$ (right), January 8--August 31, 2026. Dotted vertical lines mark March 27 and May 8; dashed horizontal lines mark the medians in the two periods before March 27 and from May 8 onward. The display of $\gamma$ is clipped at 1.5, although estimation allows values up to three.}
\label{fig:historical-parameters}
\end{figure}

\begin{figure}[htbp]
\centering
\includegraphics[width=\textwidth]{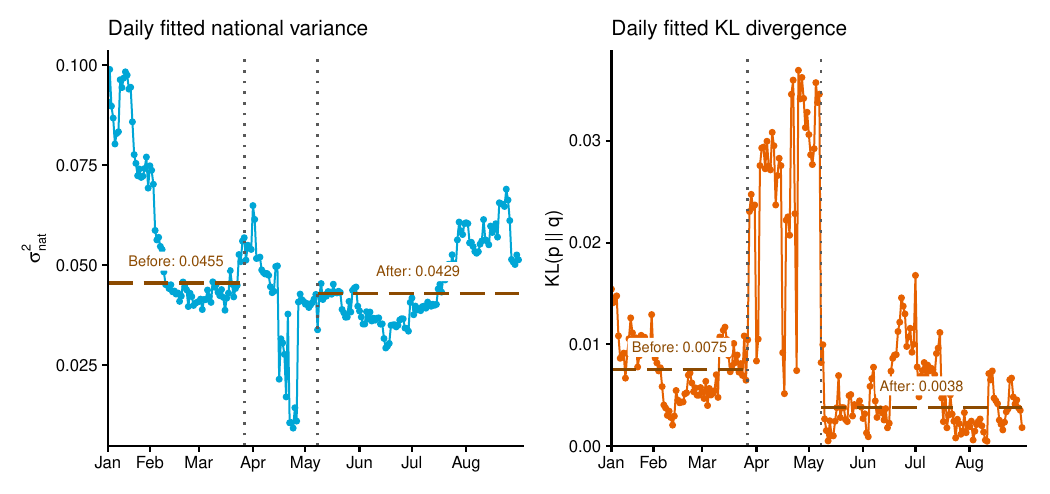}
\caption{Daily fitted national variance share $\sigma_{\mathrm{nat}}^2$ (left) and minimized forward KL divergence $D_{\mathrm{KL}}(p \parallel q)$ (right). Dotted vertical lines mark March 27 and May 8; dashed horizontal lines mark the medians in the two periods before March 27 and from May 8 onward, as in Figure~\ref{fig:historical-parameters}.}
\label{fig:historical-loss}
\end{figure}

\textbf{Parameter and Divergence Trends.}
Figure~\ref{fig:historical-parameters} displays the daily fitted $\alpha$ and $\gamma$ values, which remain relatively stable before and after the spring window but at distinctly different levels. The median of $\alpha$ is $0.052$ before March 27 and $-0.224$ from May 8 onward. Reconciliation in the earlier period therefore typically involves an upward district log-odds adjustment, whereas the later period involves a downward adjustment.

The fitted $\gamma$ values also shift between the two stable periods. The median increases from $0.529$ before March 27 to $0.758$ from May 8 onward. Because both values are below one, reconciliation in both periods involves flattening the observed market bucket probabilities, but the later period requires less flattening of these probabilities.

The latent national correlation follows a different pattern (Figure~\ref{fig:historical-loss}, left). It declines sharply from January to early February and increases noticeably during July, although the July values remain below those in January. Outside these movements and the spring volatility window, the estimates remain near similar levels in the two stable periods, with medians of $0.0455$ before March 27 and $0.0429$ from May 8 onward.

The right panel of Figure~\ref{fig:historical-loss} displays the minimized daily forward KL divergence. The divergence is generally lower during the two stable periods and rises substantially during the spring window, indicating greater difficulty in reconciling the district and aggregate markets during that period.

\textbf{Spring Volatile Period.}
The period between March 27 and May 8 represents a pronounced instability in the reconciliation of the two markets. During this window, the median forward KL divergence rises to $0.0276$, approximately $3.7$ times the median before the interval ($0.0075$) and $7.3$ times the median from May 8 onward ($0.0038$). Rather than displaying a smooth transition between the two relatively stable periods, the fitted parameters exhibit substantial day-to-day fluctuations.

The instability is especially evident in the fitted bucket exponent. The spring window contains all 36 instances in which $\gamma$ reaches either its lower or upper optimization bound over the entire historical period: twenty-eight at $0.3$ and eight at $3.0$. No $\gamma$ bound hits occur before March 27 or after May 8. Although the numerical optimization converges successfully on every date, the fitted reconciliation is highly unstable during this spring window before returning to a more stable pattern after May 8.


\section{Discussion}
\label{sec:discussion}

Our general modeling framework includes three distinct components: a correlated latent-Gaussian structure, an adjustment to input district probabilities, and an adjustment to the market bucket probabilities. Fitting the model therefore allows reconciliation through adjustments to both district and aggregate market probabilities rather than relying on district dependence alone. Such probability adjustments are motivated by pricing anomalies in prediction markets, including longshot bias, in which low-probability events may be overvalued and high-probability outcomes undervalued \citep{thaler1988parimutuel,snowberg2010explaining,page2013do}.

A related statistical issue is the potential confounding between the district log-odds multiplier $\beta$ and the bucket exponent $\gamma$. Both parameters rescale probability contrasts on a logarithmic scale, although at different levels of the model: $\beta$ rescales differences in log-odds across the marginal district probabilities, while $\gamma$ rescales pairwise log-ratios of the aggregate bucket probabilities. 
The two parameters can therefore compete to capture similar distributional adjustments during optimization. This functional overlap supports fixing $\beta=1$ and varying $\gamma$ in the primary specification, while treating the free-$\beta$ models as alternative specifications.
The snapshot results also support this parsimonious choice: the three-parameter primary model already achieves close reconciliation, with both KL divergences approximately $0.0013$, while the alternative specifications yield only modest further reductions in forward KL divergence.

The historical daily analysis also highlights the dynamic nature of market alignment. The period from March 27 to May 8 marks an observed window of substantial fluctuations in the fitted parameter and divergence values. These dates are descriptive bounds rather than formally estimated political change points. It is nevertheless noteworthy that the spring of 2026 represented a period of significant domestic and international volatility affecting U.S. elections. For instance, a two-week ceasefire with Iran was announced on April 7, alongside other related events that introduced sudden shocks into the political environment. The widely fluctuating parameter estimates, particularly the boundary behavior of $\gamma$, suggest that the district and aggregate markets may have processed these macro-level shocks differently, temporarily deviating from their established structural relationship before settling into a new alignment.

Finally, while the reconciliation results are encouraging, several limitations remain regarding their interpretation.
The fitted model minimizes structural divergence between two levels of market prices rather than producing forecasts or validated outcome probabilities. Accordingly, the estimated adjustments indicate the direction required for cross-market alignment but do not establish which market is closer to true election probabilities.
Similarly, the minimized KL divergences should be interpreted as geometric measures of distributional alignment rather than as statistical likelihoods or hypothesis tests.
Indeed, our analysis differs from the usual statistical exercise of fitting a model to a sample drawn from a well-defined population. Formal mathematical theory for this type of reconciliation approach remains to be further developed.


\section{Conclusion}
\label{sec:conclusion}

The snapshot and historical daily results demonstrate how the individual district and aggregate seat-count markets 
can be closely reconciled using a parsimonious three-parameter model. By combining a correlated latent-Gaussian structure with targeted adjustments to both district-level probabilities and aggregate bucket probabilities, we provide a mathematically concrete bridge between two distinct levels of market pricing.

Beyond its descriptive capabilities, our modeling framework offers potential practical value as a diagnostic tool for analyzing cross-market consistency. By fitting the model over time to track the relationship between individual district and aggregate markets, the estimated parameters and divergence measures can be used to identify periods of relative misalignment and to examine how different market tiers respond to macro-level shocks, as illustrated by the substantial fluctuations during the spring of 2026.

More broadly, our work develops an applied statistical framework for reconciling probabilities across different levels of aggregation in hierarchical prediction markets. Although the framework is evaluated here using prediction markets for a specific U.S. House election cycle, the underlying reconciliation problem is more general. Future research could apply and extend the framework to other hierarchically structured prediction or financial markets.




\bibliographystyle{apalike}
\bibliography{references}

\begin{thebibliography}{}

\bibitem[Albert and Chib, 1993]{albert1993bayesian}
Albert, J.~H. and Chib, S. (1993).
\newblock Bayesian analysis of binary and polychotomous response data.
\newblock {\em Journal of the American Statistical Association},
  88(422):669--679.

\bibitem[Ali and Silvey, 1966]{ali1966general}
Ali, S.~M. and Silvey, S.~D. (1966).
\newblock A general class of coefficients of divergence of one distribution
  from another.
\newblock {\em Journal of the Royal Statistical Society: Series B},
  28(1):131--142.

\bibitem[Byrd et~al., 1995]{byrd1995limited}
Byrd, R.~H., Lu, P., Nocedal, J., and Zhu, C. (1995).
\newblock A limited memory algorithm for bound constrained optimization.
\newblock {\em SIAM Journal on Scientific Computing}, 16(5):1190--1208.

\bibitem[Cartea et~al., 2015]{CarteaJaimungalPenalva2015}
Cartea, {\'{A}}., Jaimungal, S., and Penalva, J. (2015).
\newblock {\em Algorithmic and High-Frequency Trading}.
\newblock Cambridge University Press, Cambridge.
\newblock p. 18.

\bibitem[Chen and Liu, 1997]{chen1997}
Chen, S.~X. and Liu, J.~S. (1997).
\newblock Statistical applications of the {Poisson}-binomial and conditional
  {Bernoulli} distributions.
\newblock {\em Statistica Sinica}, 7(4):875--892.

\bibitem[Chen and Pennock, 2010]{chen2010designing}
Chen, Y. and Pennock, D.~M. (2010).
\newblock Designing markets for prediction.
\newblock {\em AI magazine}, 31(4):42--52.

\bibitem[Cherubini et~al., 2004]{cherubini2004copula}
Cherubini, U., Luciano, E., and Vecchiato, W. (2004).
\newblock {\em Copula Methods in Finance}.
\newblock John Wiley \& Sons.

\bibitem[Golub and Welsch, 1969]{golub1969calculation}
Golub, G.~H. and Welsch, J.~H. (1969).
\newblock Calculation of {Gauss} quadrature rules.
\newblock {\em Mathematics of Computation}, 23(106):221--230.

\bibitem[{Kalshi}, 2026a]{KalshiHouseSeats}
{Kalshi} (2026a).
\newblock Democratic house seats.
\newblock Accessed September 27, 2026.

\bibitem[{Kalshi}, 2026b]{KalshiNJ7}
{Kalshi} (2026b).
\newblock New {Jersey}'s 7th congressional district election, 2026.
\newblock Accessed September 27, 2026.

\bibitem[{Kalshi}, 2026c]{kalshiorderbook}
{Kalshi} (2026c).
\newblock Orderbook responses.
\newblock Kalshi API Documentation.
\newblock Accessed September 16, 2026.
  \url{https://docs.kalshi.com/getting_started/orderbook_responses}.

\bibitem[Page and Clemen, 2013]{page2013do}
Page, L. and Clemen, R.~T. (2013).
\newblock Do prediction markets produce well-calibrated probability forecasts?
\newblock {\em The Economic Journal}, 123(568):491--513.

\bibitem[Platt, 2000]{platt2000}
Platt, J.~C. (2000).
\newblock Probabilities for {SV} machines.
\newblock In Smola, A.~J., Bartlett, P.~L., Sch{\"o}lkopf, B., and Schuurmans,
  D., editors, {\em Advances in Large-Margin Classifiers}, pages 61--74. MIT
  Press, Cambridge, MA.

\bibitem[{Predexon}, 2026]{predexon}
{Predexon} (2026).
\newblock Prediction market data and orderbook history {API}.
\newblock Predexon Documentation.
\newblock Accessed September 16, 2026. \url{https://predexon.com/}.

\bibitem[{Python Software Foundation}, 2025]{Python313}
{Python Software Foundation} (2025).
\newblock Python.
\newblock Version 3.13.2.

\bibitem[{R Core Team}, 2026]{RCoreTeam2026}
{R Core Team} (2026).
\newblock {\em R: A Language and Environment for Statistical Computing}.
\newblock R Foundation for Statistical Computing, Vienna, Austria.
\newblock Version 4.6.1.

\bibitem[Snowberg and Wolfers, 2010]{snowberg2010explaining}
Snowberg, E. and Wolfers, J. (2010).
\newblock Explaining the favorite-long shot bias: Is it risk-love or
  misperceptions?
\newblock {\em Journal of Political Economy}, 118(4):723--746.

\bibitem[Thaler and Ziemba, 1988]{thaler1988parimutuel}
Thaler, R.~H. and Ziemba, W.~T. (1988).
\newblock Anomalies: Parimutuel betting markets: Racetracks and lotteries.
\newblock {\em Journal of Economic Perspectives}, 2(2):161--174.

\bibitem[{U.S. Census Bureau}, 2021]{censusdivisions}
{U.S. Census Bureau} (2021).
\newblock Geographic terms and definitions.
\newblock U.S. Census Bureau Website.
\newblock Accessed September 17, 2026.
  \url{https://www.census.gov/programs-surveys/popest/about/glossary/geo-terms.html}.

\bibitem[Wolfers and Zitzewitz, 2004]{wolfers2004prediction}
Wolfers, J. and Zitzewitz, E. (2004).
\newblock Prediction markets.
\newblock {\em Journal of Economic Perspectives}, 18(2):107--126.

\end{thebibliography}

\end{document}